\documentclass[12pt]{article}

\usepackage{pgfplotstable}
\usepackage[left=2.6cm,top=2.6cm,right=2.6cm,bottom=2.6cm,nohead]{geometry}
\usepackage{caption}
\usepackage{pdflscape}
\usepackage[round]{natbib}
\usepackage[makeroom]{cancel}
\usepackage{color}
\usepackage{graphicx}
\usepackage{epsfig,amsmath,latexsym,amssymb}
\usepackage{bm}
\usepackage{hyperref}
\usepackage{url}
\usepackage{booktabs}
\usepackage{multirow}
\usepackage{algorithm}
\usepackage{algpseudocode}
\usepackage{rotating}
\usepackage{tikz}
\usepackage{threeparttable}

\usepackage{comment}
\usepackage{authblk}

\makeatletter

\def\thanks#1{\protected@xdef\@thanks{\@thanks
        \protect\footnotetext{#1}}}

\makeatother

\title{\textbf{Scenario-based actuarial climate risk assessment via calibration of the DICE model to the shared socioeconomic pathways}}

\author[1]{Daisuke Murakami}
\author[2]{Pavel~V.~Shevchenko%
\thanks{Date:\ 31 October 2025}
\thanks{Address correspondence to Pavel~V.~Shevchenko; e-mail:\ pavel.shevchenko@mq.edu.au}}
\author[3]{Tomoko Matsui}
\author[4,6]{Aleksandar~Arandjelovi\'{c}}
\author[5]{Tor~A.~Myrvoll}

\affil[1]{\footnotesize Department of Fundamental Statistical Mathematics, Institute of Statistical Mathematics, Japan}
\affil[2]{\footnotesize Department of Actuarial Studies and Business Analytics, Macquarie University, Australia}
\affil[3]{\footnotesize Department of Interdisciplinary Statistical Mathematics, Institute of Statistical Mathematics, Japan}
\affil[4]{\footnotesize Institute for Statistics and Mathematics, Vienna University of Economics and Business, Austria}
\affil[5]{\footnotesize Department of Electronic Systems, Norwegian University of Science and Technology, Norway}
\affil[6]{\footnotesize Department of Mathematics, ETH Zurich, Switzerland}

\date{}

\begin{document}

\maketitle
\thispagestyle{empty}

\vspace{-36.5pt}

\begin{abstract}
\noindent
Accounting for climate-related risks is an emerging problem for life insurers around the world. In this paper, we demonstrate how scenario trajectories for global temperature can be obtained using the cost-benefit Dynamic Integrated Climate-Economy (DICE) model calibrated to the five Shared Socioeconomic Pathways (SSPs). These scenarios can also be calculated under different carbon emission mitigation targets such as achieving net-zero carbon emissions by a specific year.
We show how to calibrate the DICE model to align industrial and land-use carbon emissions with projections from six leading process-based integrated assessment models (IAMs): IMAGE, MESSAGE--GLOBIOM, AIM/CGE, GCAM, REMIND--MAgPIE and WITCH--GLOBIOM.
The obtained scenario trajectories of global temperature can be linked to the climate-change induced excess mortality in various regions that, in turn, can be used for stress testing of life insurance portfolios. We illustrate this using synthetic portfolios of life insurance and annuity products.

\medskip\noindent\textbf{Keywords:} Actuarial climate risk, climate scenarios, life insurance, integrated climate-economy modelling, shared socioeconomics pathways, net-zero emission scenarios
\end{abstract}





\clearpage

\section{Introduction}\label{sec:introduction}
Global warming prompted the negotiation of the Paris Agreement in 2015 by 196 parties \cite[][]{ParisAgreement2015} aiming to limit the global temperature increase to $2^\circ\text{C}$ above preindustrial levels, with a preferred target of $1.5^\circ\text{C}$. To achieve these goals, many countries, cities, businesses, and other institutions have committed to achieving net-zero carbon emissions.
Hereafter, ``carbon emission'' refers to the emission of all greenhouse gases.
Over 140 countries, including China, India, and the European Union countries, have set goals to reach net-zero emissions by around 2050-2070.
This has resulted in more public and private funds being provided for green projects involving renewable energy and energy conservation, as the world works to mitigate the effects of global climate change.

The impact of climate change on life and non-life insurance portfolios becomes an important area of risk management. Climate change increases not only the frequency and severity of weather-related non-life insurance claims but also mortality rates affecting life insurance products.  Stress testing is a common practice of risk management undertaken by insurance companies, but is little studied with respect to climate change in the academic literature. Our paper demonstrates how  stress test scenarios for temperature trajectories can be obtained and linked to temperature-related excess mortality for stress testing of life insurance portfolios.

A number of actuarial societies have developed the so-called climate indices\footnote{See for example, The Actuaries Climate Indexes for USA and Canada, and Australia \url{https://actuariesclimateindex.org/} and \url{https://www.actuaries.asn.au/microsites/climate-index/} }
measuring the observed changes in
extreme weather and sea levels to inform actuaries about climate trends and impacts
related to non-life insurance. At the same time, climate change can significantly impact mortality rates by exacerbating heat-related, cardiovascular and respiratory illnesses. Of course, the impact of climate change on mortality depends on geographic location and death rate  may even decrease in some cold
climate countries. However, overall the increase in mortality on every continent is confirmed by several studies, see e.g. \cite{vicedo2021burden, gasparrini2017projections}.

The European Insurance and Occupational Pensions Authority (EIOPA) put forward stress testing principles for insurers \cite[][]{EIOPA2022methodological} that provides high level guidance to insurers. The most recent application paper by the International Association of Insurance Supervisors  \cite[][]{iais2025application} discusses the key topics relevant for the supervision of climate-related risks including stress test/scenario analysis. It defines climate-related risk (page 10) as ``\emph{risk posed by the exposure of an insurer to physical, transition
and/or litigation risks caused by or related to climate change.}"
It also defines (page 75) scenario analysis as ``\emph{A method of assessment that considers the impact of a combination of circumstances to reflect historical or other scenarios which are analysed in light of current conditions. Such analysis may be conducted deterministically or stochastically}". It recommends to consider
a very wide range scenarios for global emissions including net-zero scenarios. Although it does not prescribe specific scenarios, it mentions scenarios developed by the Network of Central Banks and Supervisors for Greening the Financial System (NGFS)\footnote{\url{https://www.ngfs.net/}}, see \cite{NGFS2024}. These scenarios are based on several process-based Integrated Assessment Models (IAMs) and Shared Socioeconomic Pathways (SSPs).

Since the early 1990s, IAMs have become a central analytical tool in climate change economics; see reviews in \cite{weyant2017some}.
The mitigation and climate impact assessments in the Intergovernmental Panel on Climate Change (IPCC) Assessment Reports\footnote{The IPCC is the United Nations body responsible for assessing the science related to climate change. Its latest report, the Sixth Assessment Report (2023), is available at \url{www.ipcc.ch}.} have primarily relied on IAMs developed by various research groups.
These models can be broadly categorised into two types: cost-benefit IAMs, which estimate optimal mitigation strategies by balancing economic costs and climate damages, and process-based IAMs, which form the basis of IPCC's assessments of transformation pathways towards temperature targets.
However,  cost-benefit IAMs have historically relied on a single reference optimal pathway and have not systematically incorporated the full range of Shared Socioeconomic Pathway (SSP) scenarios.
As a result, there is a lack of integrated research pursued in our paper that quantitatively aligns the baseline assumptions of process-based IAMs with cost-benefit modelling frameworks under alternative net-zero emission trajectories.

The climate change research community has developed five socioeconomic narratives, known as Shared Socioeconomic Pathways (SSPs): SSP1 -- SSP5.
These scenarios cover a broad spectrum of future challenges to mitigation and adaptation, offering projections on key factors such as demographics, economic growth, energy use, land use, and air pollution.
The SSP framework was initially proposed by \cite{moss2010next} and
\cite{van2012proposal}, with its quantified version published several years later \cite{riahi2017shared}.
It serves as a foundation for integrated analyses of climate mitigation and adaptation strategies.
SSPs are also used as inputs for the latest climate models, with their outputs informing the  IPCC assessment reports.

In addition, researchers have developed the Representative Concentration Pathways (RCPs), which describe different trajectories of greenhouse gas concentrations leading to specific levels of radiative forcing (measured in watts per square meter) by 2100.
The SSP and RCP frameworks are designed to be complementary: RCPs define pathways for greenhouse gas concentrations, while SSPs provide socioeconomic contexts that influence emission reductions.
Together, an SSP-RCP combination represents a possible future scenario.
The \textbf{SSP baseline} scenario assumes the absence of any concerted international effort to address climate change, beyond those already adopted by countries.

In this paper, we develop scenarios of global temperature and use these scenarios to find projections of mortality rates driven by changes in the temperature that affect life insurance products.
We study this using the cost-benefit dynamic integrated climate-economy (\textbf{DICE}) model calibrated to baseline SSPs and emissions from the leading process-based models. DICE was originally proposed more than 30 years ago in \cite{nordhaus1992dice} and was regularly refined with the latest beta version update in 2023\footnote{The most recent version of the DICE model is available at {\url{https://williamnordhaus.com}}.}. It is a cost-benefit IAM that estimates optimal mitigation levels relative to the economic costs of
climate impacts.
DICE model is one of the three main IAMs used by the United States government to determine the social cost of carbon; see
\cite{scc2016technical}. The other two are PAGE (Policy Analysis of the Greenhouse Effect
 model, see \cite{hope2008optimal}) and FUND (Climate Framework
for Uncertainty, Negotiation, and Distribution model, see \cite{anthoff2010}).
Even though the DICE model has limitations in the model structure and model parameters
debated in the literature \cite[][]{grubb2021modeling,pindyck2020use}, it has become the typical reference point
for climate-economy modelling, and is used in our study. The DICE model is a global model but fits for purpose with respect to modelling global temperature and global carbon concentration, though it may need  recalibration of model parameters in view of recent findings in \cite{folini2025climate}. We also note that, regional version of DICE (RICE model) is available \cite[][]{nordhaus1996regional} that allows for multiple regions to model region specific economies but its climate model is the same as in DICE.

In our study, to calibrate DICE emissions under the baseline SSPs, we use all process-based IAMs available in the SSP database hosted in the International Institute for Applied Systems Analysis (IIASA)\footnote{https://tntcat.iiasa.ac.at/SspDb} hereafter referred to as SSPdata-IIASA. These six IAMs are: \textbf{IMAGE} (Integrated Model to Assess the Global Environment), \textbf{MESSAGE-GLOBIOM} (Model for Energy Supply Systems And their General Environmental Impact -- Global Biosphere Management Model), \textbf{AIM/CGE} (Asia-Pacific Integrated Model/ Computable General Equilibrium), \textbf{GCAM} (Global Change Analysis Model) and \textbf{REMIND-MAgPIE} (Regional Model of Investment and Development -- Model of Agricultiral Production and its Impact on the Environment), and \textbf{WITCH-GLOBIOM} (World Induced Technical Change Hybrid -- Global Biosphere Management Model).  See \cite{van2020anticipating} for a description of the historical context of IAMs and their growing popularity in the climate-economy literature. 

Almost any process-based IAM can be run for different SSPs. Among these, for each SSP a single IAM interpretation was selected as the so-called representative marker scenario for recommended use by future analyses of climate change, its impacts, and response measures. These are referred to as \textbf{marker SSPs}, discussed in detail in e.g. \cite{riahi2017shared} and defined in Section \ref{DICE_calibration_sec}. For example, marker SSP1 corresponds to SSP1 run using IMAGE.
The selection of markers was guided by two main considerations: the internal consistency of the full set of SSP markers, and the ability of the different models to represent distinct characteristics of the storylines.
 The SSP data and process-based IAMs forecasts are available from SSPdata-IIASA. This database does not include cost-benefit IAMs such as DICE. Even though marker SSPs are standard references, non-marker SSPs are also considered to be important since they provide insights into possible alternative scenario interpretations of the same basic SSP
elements and storylines. In our study, we calculate results for DICE model calibrated to baseline scenarios of all five marker SSPs; and we also calculate non-marker scenarios when IAM data are available for all five SSPs.

In our study, we re-calibrate the DICE model to match the SSP baseline marker and non-marker scenarios\footnote{The baseline SSP scenarios are the reference cases for mitigation, climate impacts and
adaptation analyses. They represent future developments in the absence of
new climate policies and mitigation scenarios.} under the main six process-based IAMs for the emission of greenhouse gases. Then, re-calibrated DICE model is used to produce scenarios of future for climate and economy variables under SSP1--SSP5 and assuming scenarios for carbon emission reduction leading to net-zero carbon emission in 2050 and 2100 (we also calculated scenarios for zero industrial emission only). One of the main observations from these projections is that even if a zero emission target is achieved worldwide by 2050, the temperature increased by 2100 will be larger than 2$^\circ$C in the range 2.5-2.7$^\circ$C. For this scenario, the social cost of carbon is increasing about 10 times from USD 30-50 in 2025 to USD 250-400 in 2100 depending on SSP-IAM. This provides a carbon price benchmark for policy makers and is generally consistent with \cite{riahi2022mitigation} and  \citet{yang2018social}.

If temperature-related excess mortality is estimated, then  temperature projections translate into scenarios for mortality rates, that in turn impact insurance premia and reserves. In this paper,  we use climate-induced mortality from \cite{bressler2021mortality}.  Note that for illustration we use global temperature related excess mortality function, but for practical use, one should use country specific excess mortality functions such as estimated in \cite{lee2020projections} or \cite{gasparrini2017projections}.
The scenarios based on the DICE model developed in our paper should be considered as complementing NGFS scenarios developed using three process-based models (REMIND-MAgPIE, MESSAGE-GLOBIOM, and GCAM). The cost-benefit DICE model maximises the welfare function wrt consumption and emission controls, i.e. allows for adaptive optimal decisions rather than fixed scenarios under the process-based models.


The remainder of this paper proceeds as follows. The model and methodology are defined in Section~\ref{sec:model}.  The results are presented in Section~\ref{sec:results}, and some concluding remarks are given in Section~\ref{sec:conclusion}.

\section{Methodology}\label{sec:model}
In this section, we outline the main DICE model equations,  the calibration of the DICE exogenous functions to the SSP-IAM data, and the excess temperature-related mortality function. The results for climate-economy trajectories and the impoacts on life insurance products under DICE calibrated to different SSPs-IAMs will be presented in the next section.
\subsection{Model}
In our study, we calculate the climate-economy projections using DICE-2016. Specifically, we use the version {\tt DICE2016R-091916ap.gms}. 
We note that beta 2023 version is available, but we prefer to use the 2016 version, which is well tested and analysed in various studies. For details of DICE, we refer to \cite{nordhaus2017revisiting} and the references therein. This model maximises the social welfare function
\begin{equation}\label{welfare_fnc}
W=\sum_{t=0}^\infty U(C_t,L_t) \rho^t,
\end{equation}
where $C_t$ is the consumption and  $L_t$ is the population of the world in billions at time $t$,
$U(C_t,L_t)=L_t(C_t/L_t)^{1-\alpha}/(1-\alpha)$ is the utility per period, $\alpha=1.45$ is the risk aversion parameter, $\rho$ is a utility discounting factor; and time index $t=0,1,...$ corresponds to $\Delta=5$-year time step.
This deterministic DICE-2016 model features six state variables: world produced economic capital $K_t$  in trillions of USD as of 2010, carbon concentrations (in billions of tons) in the atmosphere, the upper oceans, and the lower oceans  $M_{t}^{\mathrm{AT}}$, $M_{t}^{\mathrm{UP}}$ and $M_{t}^{\mathrm{LO}}$ respectively, the global mean surface temperature $T_{t}^{\mathrm{AT}}$, and the
mean temperature of the deep oceans $T_{t}^{\mathrm{LO}}$ (both temperatures are measured in degrees of $^\circ$C above the temperature during the year 1900).

The DICE model also features the carbon emission control $\mu_t\ge 0$ (industrial carbon emission reduction rate) for each $t$ defined such that the annual industrial $\mathrm{CO}_2$ emission is given by
\begin{equation}\label{carbon_emission_eq}
E_t^{\mathrm{Ind}}=(1-\mu_t)\sigma_t Y_t,
\end{equation}
where $Y_t$ is the gross annual economy output and $\sigma_t$ is the carbon intensity
or the $\mathrm{CO}_2$-output ratio. The total emission of $\mathrm{CO}_2$ (in billions of tons per year) is $E_t=E_t^{\mathrm{Ind}}+E_t^{\mathrm{Land}}$, where
$E_t^{\mathrm{Land}}$ is the exogenous deterministic function of land emission.

The evolution of economic capital is given by
\begin{equation}\label{economic_capital_eq}
    K_{t+1}=K_t(1-\delta_K)^\Delta+\Delta \times (Q_t-C_t),
\end{equation}
where $\delta_K$ is the depreciation parameter and
$ Q_t=\Omega_t Y_t$ is the net economic annual output in a period $t$. The gross output $Y_t$ is modelled by the Cobb–Douglas production function of capital,
labor, and technology
\begin{equation}\label{grossoutput_eq}
Y_t= A_t K_t^\gamma L_t^{1-\gamma},
\end{equation}
where $\gamma=0.3$ and $A_t$ is the total factor productivity  representing
technological progress and efficiency improvements over time. The damage-abatement factor is
\begin{equation}\label{damage_abatement_factor_eq}
\Omega_t=1-\theta_{1,t}\mu_t^{\theta_2} - \pi_2 \times (T^{\mathrm{AT}}_t)^{2},
\end{equation}
where the second term is the abatement cost and the third term is the damage cost both as the fraction of the output; $\theta_2=2.6$, $\pi_2=0.00236$ and $\theta_{1,t}$ are parameters specified in DICE.
The evolutions of carbon concentration $M_t=(M_t^{\mathrm{AT}},M_t^{\mathrm{UP}},M_t^{\mathrm{LO}})^\top$ and global temperature $T_t=(T_t^{\mathrm{AT}},T_t^{\mathrm{LO}})^\top$ are modelled as
\begin{align}\label{carbon_concentration_eq}
M_{t+1}&=\Phi_M M_t+\Delta\times(\beta E_t,0,0)^\top,\\
\label{temperature_eq}
T_{t+1}&=\Phi_T T_t+(\xi_1 F_{t+1},0)^\top,
\end{align}
where $F_t=\eta \log_2 ({M_t^{\mathrm{AT}}}/{{M}_\ast^{\mathrm{AT}}})+F_t^{\mathrm{EX}}$ is the radiative forcing, ${M}_\ast^{\mathrm{AT}}=588$ is the pre-industrial carbon concentration,  and $F_t^{\mathrm{EX}}$ is the external forcing function. $\Phi_M$ is a 3 by 3 matrix of coefficients for the evolution of concentrations and $\Phi_T$ is a 2 by 2 matrix of coefficients for the evolution of temperatures.
Here, we also note that $\beta=3.666$ represents the $\mathrm{CO}_{2}$ to carbon mass transformation coefficient, i.e., under the model parameterisation, emission is measured in tons of carbon dioxide while concentration is in tons of carbon.
For a detailed discussion and all parameter values of the DICE model, we refer to \cite{nordhaus2018projections} and the GAMS code of DICE-2016.
 In the original DICE-2016, functions $L_t$, $A_t$, $\sigma_t$, $E^{\mathrm{Land}}$ and $F_t^{\mathrm{EX}}$ are exogenous functions estimated using various data sources (see \cite{nordhaus2018projections}), different from the SSP-IAM data.

In this paper we follow \cite{bressler2021mortality}  extending the DICE model treating population as an additional state variable that also depends on temperature:
\begin{align}\label{population_eq}
L_{t+1}&=L_t\times \left(1+b_t-d_t(1+\delta(T_t^{\mathrm{AT}}))\right)\nonumber \\
&=L_t\times (1+b_t-d_t)-N_t\times \delta(T^{\mathrm{AT}}_t),
\end{align}
where $b_t$ is the fertility rate and $d_t$ is the mortality rate.
For numerical illustration, we will use the global excess mortality function $\delta(x)=a x^\nu$ with $a=0.0001811$ and $\nu=3.745$ estimated in \cite{bressler2021mortality}.  The factor $(1-b_t-d_t)$ can be calibrated using SSP data  as $(1-b_t-d_t)=\hat{L}_{t+1}/{\hat{L}_t}$, where $\hat{L}_t$ is population under the corresponding SSP, and the number of deaths $N_t=L_t d_t$ is estimated as the number of deaths projected under different SSPs in \cite[Figure 2]{sellers2020cause}.

The DICE model finds optimal carbon emission control $\mu_t$ and optimal consumption $C_t$ for each $t$ maximising the welfare function (\ref{welfare_fnc}) with an infinite time horizon replaced by $N=100$:
\begin{equation}\label{value_fnc}
V_0(X_0)=\sup_{\bm{C},\bm{\mu}}\sum_{t=0}^N U(C_t,L_t) \rho^{t},
\end{equation}
where $\bm{C}=(C_0,C_1,...,C_N)$ and $\bm{\mu}=(\mu_0,\mu_1,...,\mu_N)$ are controls and $X_t$ is the vector of state variables $X_t=(K_t,M_t^{\mathrm{AT}},M_t^{\mathrm{UP}},M_t^{\mathrm{LO}},T_t^{\mathrm{AT}},T_t^{\mathrm{LO}}, L_t)^\top$ evolving in time according to equations (\ref{economic_capital_eq},\ref{carbon_concentration_eq},\ref{temperature_eq}) that can be written as a deterministic Markov decision process $X_{t+1}=\mathcal{T}_t(X_t,C_t,\mu_t)$. This is a standard optimal control problem that can be solved using the Bellman equation backward in time for $t=N,N-1,...,0$ \cite[see e.g.][]{bauerle2011markov}
$$
V_t(X_{t})=\sup_{C_t,\mu_t}(U(C_t,L_t)+\rho V_{t+1}(X_{t+1})),\; V_{N+1}=0
$$
with optimal strategy calculated as
$$
(C^\ast_t(X_t),\mu^\ast_t(X_t))=\arg\sup_{C_t,\mu_t}(U(C_t,L_t)+\rho V_{t+1}(X_{t+1})).
$$
The orignal DICE solution in GAMS and Excel is brute force maximisation in (\ref{value_fnc}) with respect to $C_0,...,C_N$ and $\mu_0,...,\mu_N$.

The social cost of carbon (SCC), representing economic damage  expressed in dollars per additional ton of $\mathrm{CO}_{2}$ can be defined under the DICE model as
\begin{equation}\label{equation-SCC}
SCC_{t} = -1000 \beta \frac{\partial V_{t} /\partial M_{t}^{\mathrm{AT}}} {\partial V_{t} / \partial K_{t}}
\;\mbox{or}\;
SCC_{t} = -1000 \frac{\partial W /\partial E_{t}} {\partial W / \partial C_{t}}.
\end{equation}
The second formula is used in the original GAMS code implementation of DICE with derivatives calculated numerically using small perturbation of the arguments in the welfare function from their optimal values; this second representation is somewhat confusing in the original DICE literature, and we refer the reader to \cite{braun2024social} for precise mathematical formulation details. In our paper, we use the R-code implementation of DICE producing identical results to the GAMS code subject to computational precision\footnote{Our code is based on adapting the R-code  \url{https://github.com/olugovoy/climatedice}.}. For computational efficiency, we calculate $SCC_t$  directly as the net present value of damages by calculating the total discounted economic damages of the additional ton of $\mathrm{CO}_2$ released in year $t$ (the additional ton of $\mathrm{CO}_2$ yields reduced economic consumption in the future as the result of a
climate damage function). This gives virtually the same results for SCC as obtained by the original GAMS code \cite[p.284][]{nordhaus2014estimates}; see also \cite{braun2024social} for the numerical calculation of SCC using different methods.

In our paper,
we also consider deterministic scenarios for $\mu_t$, that requires maximisation of the welfare function with respect to consumption only. Given scenarios for $\mu_t$ and optimal consumption values of $C_t$, the trajectories of the economic and state variables are calculated using  (\ref{economic_capital_eq},\ref{carbon_concentration_eq},\ref{temperature_eq}).

Note that $\mu_t=1$ corresponds to zero industrial carbon emission, i.e., $E_t^{\mathrm{Ind}}=0$. However, net-zero targets correspond to zero of the total carbon emission $E_t^{\mathrm{Ind}}+E_t^{\mathrm{Land}}$. To consider net-zero scenarios we re-parameterise the model to have emission control $\widetilde\mu_t$ so that the total emission is
$$
E_t=(1-\widetilde\mu_t)\times\left(\sigma_t Y_t+ E_t^{\mathrm{Land}}\right),
$$
i.e. $E_t=0$, when $\widetilde\mu_t=1$. The original emission control under this parameterisation is calculated as $\mu_t=\widetilde\mu_t\times
\left(1+E_t^{\mathrm{Land}}/(\sigma_t Y_t)\right)$.

\subsection{Climate change induced excess mortality}
Mortality rate projections is the main input for life insurance products and reserves. Various deterministic and stochastic mortality models have been studied in the literature \cite[][]{booth2008mortality}.
Denote the baseline annual probability of death at age $x$ in year $\tau$ by $q_x(\tau)$, then the mortality rate accounting for the temperature-related excess mortality can be written as
$$
q^\mathrm{adj}_x(\tau)=q_x(\tau)\times \big(1+\widetilde\delta(T^\mathrm{AT}(\tau))\big).
$$
Here, $\widetilde\delta(\cdot)$ is temperature related excess mortality function, where $T^\mathrm{AT}(\tau)$ denotes the global atmospheric temperature in year $\tau$. Note that time period index $t$ in the DICE-2016 model corresponds to a five-year time step and interpolation is needed to estimate $T^\mathrm{AT}$ in year $\tau$.

We also remark that excess mortality $\widetilde\delta(\cdot)$ can depend on age $x$ but these data are not available to us and for numerical examples we use the age independent function $\delta(\cdot)$ estimated in \cite{bressler2021mortality}. Then the survival probability over $k$ years of an individual aged $x$ can be calculated as
\begin{equation}
{_k}p_x^{\mathrm{adj}}(\tau)=\prod_{j=0}^{k-1}(1-q_{x+j}^\mathrm{adj}(\tau+j)).
\end{equation}

Once scenario realisation of $T_t^\mathrm{AT}$ is known for $t=0,1,...$, then $\widetilde\delta(T^\mathrm{AT}(\tau))$ can easily be evaluated for each future year $\tau$.
Practitioners can use available temperature projections from various process based IAMs or/and NGFS scenarios. In our paper, we develop scenarios based on the cost-benefit DICE model that maximises the welfare function wrt consumption and emission controls, i.e. allows for adaptive optimal decisions rather than fixed scenarios under the process-based models.

Finally, we note that the excess mortality function $\widetilde\delta(\cdot)$ for the purposes of the evaluation of life insurance products and portfolios should be country specific or even portfolio specific rather than global excess mortality function $\delta(T^\mathrm{AT}_t)$ in the world population dynamics of $L_t$ in (\ref{population_eq}).

\subsection{DICE calibration}\label{DICE_calibration_sec}
Population $L_t$, total factor productivity $A_t$, land emission $E_t^{\mathrm{Land}}$, and decorbanisation $\sigma_t$ are exogenous parametric  functions specified in DICE. In the original DICE, these were estimated using various data sources, resulting in a single scenario of the future for the economy and climate. In our study we estimate these functions from population, GDP and emission data of the baseline SSPs and process-based IAMs, to get many plausible scenarios to assess the possible trajectories of climate and economic variables in the future. Also note that, in our extended DICE model, $L_t$ is an endogenous state variable driven by global atmospheric temperature (\ref{population_eq}).

The SSPs provide quantitative projections and qualitative
descriptions. Quantitative projections include population and economic growth. A detailed description of SSPs can be found in \cite{riahi2017shared}. In short, they are referred as follows.
\begin{itemize}
\item \textbf{SSP1 -- ``Sustainability - Taking the Green Road"} (low challenges to mitigation and adaptation), representing the sustainable development.
\item \textbf{SSP2 -- ``Middle of the Road"} (medium challenges to mitigation and adaptation), implying a development
pathway consistent with typical historical patterns.
\item \textbf{SSP3 -- ``Regional Rivalry – A Rocky Road"} (high challenges to mitigation and adaptation), characterised by international fragmentation and regional
rivalry.
\item \textbf{SSP4 -- ``Inequality – A Road Divided (low challenges to mitigation, high challenges to adaptation)''}, characterised by extreme inequality both across and within countries.
\item \textbf{SSP5 -- ``Fossil-fueled Development – Taking the Highway"} (high challenges to mitigation, low challenges to adaptation), forecasting economic successes for both industrialized and emerging economies.
\end{itemize}

The SSP baseline scenarios provide a description of future developments in the absence of new climate policies beyond those in place today, as well as mitigation scenarios which explore the implications of climate change mitigation policies. The baseline SSP scenarios are reference cases for mitigation, climate impacts, and
adaptation analyses.

Many process-based IAMs have been developed in climate change economics. These are detailed, multi-region, large-scale computer simulation models to assess policy options for stabilising the global climate. They couple representations of the economy, the energy system, the agricultural and land use system, and the climate system. The specific IAMs considered in our study will be described below.
In principle, each process-based IAM can be run for each SSP. However, for reference, for each SSP a single IAM interpretation was selected as the so-called representative marker scenario for recommended use by future analyses of climate change, its impacts, and response measures. These are the so-called \emph{marker SSPs} selected as representative of the broader developments of each SSP, guided by the internal consistency and the ability of the different models to represent
distinct characteristics of the storylines.
Below is a summary list of marker SSPs:
\begin{itemize}
\item \textbf{marker SSP1}: SSP1 with IMAGE \cite[][]{van2017energy};
\item \textbf{marker SSP2}: SSP2 with MESSAGE-GLOBIOM \cite[][]{fricko2017marker};
\item \textbf{marker SSP3}: SSP3 with AIM/CGE \cite[][]{fujimori2017ssp3};
\item \textbf{marker SSP4}: SSP4 with GCAM \cite[][]{calvin2017ssp4};
\item \textbf{marker SSP5}: SSP5 with REMIND-MAgPIE \cite[][]{kriegler2017fossil}.

\end{itemize}

\begin{figure}[!h]
  \centering
  \includegraphics[width=17cm]{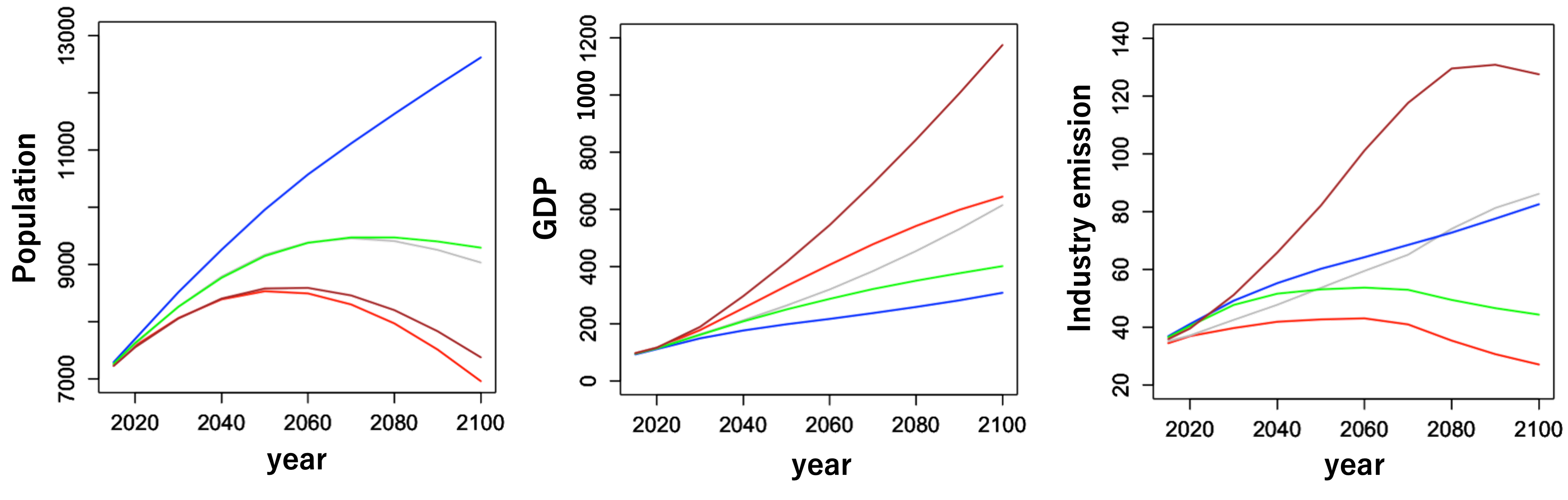}
  \caption{Global population (millions), GDP (trillions 2005 USD per year), and industrial $\mathrm{CO}_2$ emission (Gigatons) scenarios by marker SSPs (Red: SSP1, Grey: SSP2, Blue: SSP3, Green: SSP4, Brown: SSP5).}\label{fig_ssp}
\end{figure}

For a detailed discussion, see also \cite{riahi2017shared}. Figure \ref{fig_ssp} shows the data for the world GDP, population, and emission under five marker SSPs (baselines). Here, we note that only under SSP3, the world population keeps growing from now to 2100. Under other SSPs, the population peaks at around 2050-2070. It is the lowest under SSP1 and the largest under SSP3. GDP grows over time under all SSPs; it is the lowest under SSP3 and the largest under SSP5. Industry emission grows under SSP2 and SSP3. It increases until 2050-2060 for SSP1 and SSP4, and then decreases. Under SSP5, emission is the largest, growing fast until around 2080 where it flattens. Under SSP1, emission is the lowest.

The SSP-IAM data are available from the SSPdata-IIASA database.
In addition to data from marker SSPs, we also considered data for non-marker SSPs. Not all IAMs can run for all SSPs. For non-marker scenarios, we consider four IAMs: AIM/CGE, GCAM, IMAGE and WITCH-GLOBIOM \cite[][]{emmerling2016witch}
that have results under all five SSPs.


 We use the baseline SSP scenarios from the process-based IAMs to estimate the DICE exogenous functions as follows\footnote{This approach is motivated by paper paper \citet{yang2018social} that examines the DICE results
under the marker SSP1–SSP5, where the results of the China Climate Change integrated
assessment model (C3IAM) were used to re-estimate some parameters in the DICE model.} (the corresponding variables taken from the SSP-IAM scenario are indicated by $\widehat{}$ ).

\begin{itemize}
\item The population $L_t$ is taken from the SSP-IAM scenario, i.e. $L_t=\widehat{L}_t$. Here we remark that, we tested to use $L_t$ as endogenous variable evolving in time according to (\ref{population_eq}), where death and fertility rates are estimated from SSP data but the impact of the excess mortality function $\delta(T^{\mathrm{AT}}_t)$ on the temperature projections and other DICE outputs appeared to be not material, and thus one can simply use $L_t=\widehat{L}_t$.

\item The decarbonisation function $\sigma_t$ is estimated as $\sigma_t=\widehat{E}^{\mathrm{Ind}}_t$/$\widehat{Y}_t$, where $\widehat{E}^{\mathrm{Ind}}_{t}$ is taken from the IAM baseline scenarios and $\widehat{Y}_t$ is the gross GDP in the baseline SSPs.
\item Land emission $E^{\mathrm{Land}}_t$ is taken from SSPs, i.e., $E^{\mathrm{Land}}_t=\widehat{E}^{\mathrm{Land}}_t$.
\item The total productivity factor $A_t$ is calibrated so that the gross GDP $Y_t$ from DICE matches $\widehat{Y}_t$ the gross GDP in the baseline SSPs. This calibration involves iterative process 
that, for each iteration $m=1,2,...$ calculates the DICE trajectories of $Y^{(m)}_t$ and model state variables $X^{(m)}_t$, $t=0,1,...,N$ using
\begin{equation}\label{iterative_solution_eq}
A^{(m)}_t=\frac{\widehat{Y}_t}{\big(K^{(m-1)}_t\big)^\gamma \widehat{L}_t^{1-\gamma}},\; t=0,1,...,N
\end{equation}
until trajectory of $Y^{(m)}_t$  approximates the SSP trajectory of $\widehat{Y}_t$ closely. $K_t^{(m-1)}$, $t=0,1,...,N$ is calculated according to (\ref{economic_capital_eq}) and (\ref{grossoutput_eq}) via $A^{(m-1)}_t$, $t=0,1,...,N$, i.e. iterative process (\ref{iterative_solution_eq}) is the standard fixed-point iteration method. Here we note that: each iteration involves solving the DICE model (optimising the welfare function  with respect to the consumption) without climate damage and abatement costs; $K^{(0)}_t$ is taken from the original DICE solution; and convergence is achieved quickly after few iterations.
\end{itemize}

Then for each baseline SSP and process-based IAM we solve DICE (miximising the welfare function with respect to consumption $C_t$) resulting in determinstic trajectory for temperature, carbon concentration, and social cost of carbon. We do not optimise welfare function in DICE for emission control $\widetilde\mu_t\in[0,1]$ but we consider scenarios when net-zero emission is achieved by a specific year as follows:
\begin{itemize}
    \item Net-zero in 2050: $\widetilde\mu_t$ increases linearly from 0 in 2015 to 1 in the year 2050 and remains 1 afterward; and
    \item Net-zero in 2100: $\widetilde\mu_t$ increases linearly from 0 in 2015 to 1 in the year 2100 and remains 1 afterward.
\end{itemize}
We also consider emission control scenarios of achieving zero industrial emission:
\begin{itemize}
    \item zero industrial emission in 2050: $\mu_t$  increases linearly from 0 in 2015 to 1 in 2050 and remains 1 afterward; and
    \item zero industrial emission in 2100: $\mu_t$ is increases linearly from 0 in 2015 to 1 in year 2100 and remains 1 afterward.
\end{itemize}

For comparison, under the original DICE-2016 model, optimal $\mu_t$ reaches 1 in 2115.
The idea of considering scenarios for greenhouse gas emission $\mu_t$ under the DICE framework is not new. For example, it was previously attempted under DICE-2008 in \cite{howarth2014risk} and \cite{gerst2013interplay}, where the emission control rate scenario rises from a value of zero in 2010 to unity in the year 2270.

It is important to note that GDP in the SSP database is in 2005 USD year rate while the DICE model assumes 2010 year rate. The SSP's GDP is converted to the 2010 year rate by multiplying 1.14 that is the corresponding growth rate of the consumer price index (CPI) evaluated by the CPI Inflation Calculator from the US Bureau of Labor Statistics (\url{https://www.bls.gov/data/inflation_calculator.htm}). Finally, since SSPs are available only until 2100, to run the DICE model, we project the SSP-IAM population, GDP, and carbon emission scenarios after 2100 using log-linear trend extrapolation method, e.g. see \cite{yamagata2015comparison}.



\section{Results}
\label{sec:results}
In this section, firstly we show results for climate-economy projections and then show impact of temperature-related excess mortality on two synthetic life portfolios of a life insurer.

\subsection{DICE projections under SSPs and net-zero scenarios}



\begin{figure}[h!]
  \centering
  \includegraphics[width=15cm]{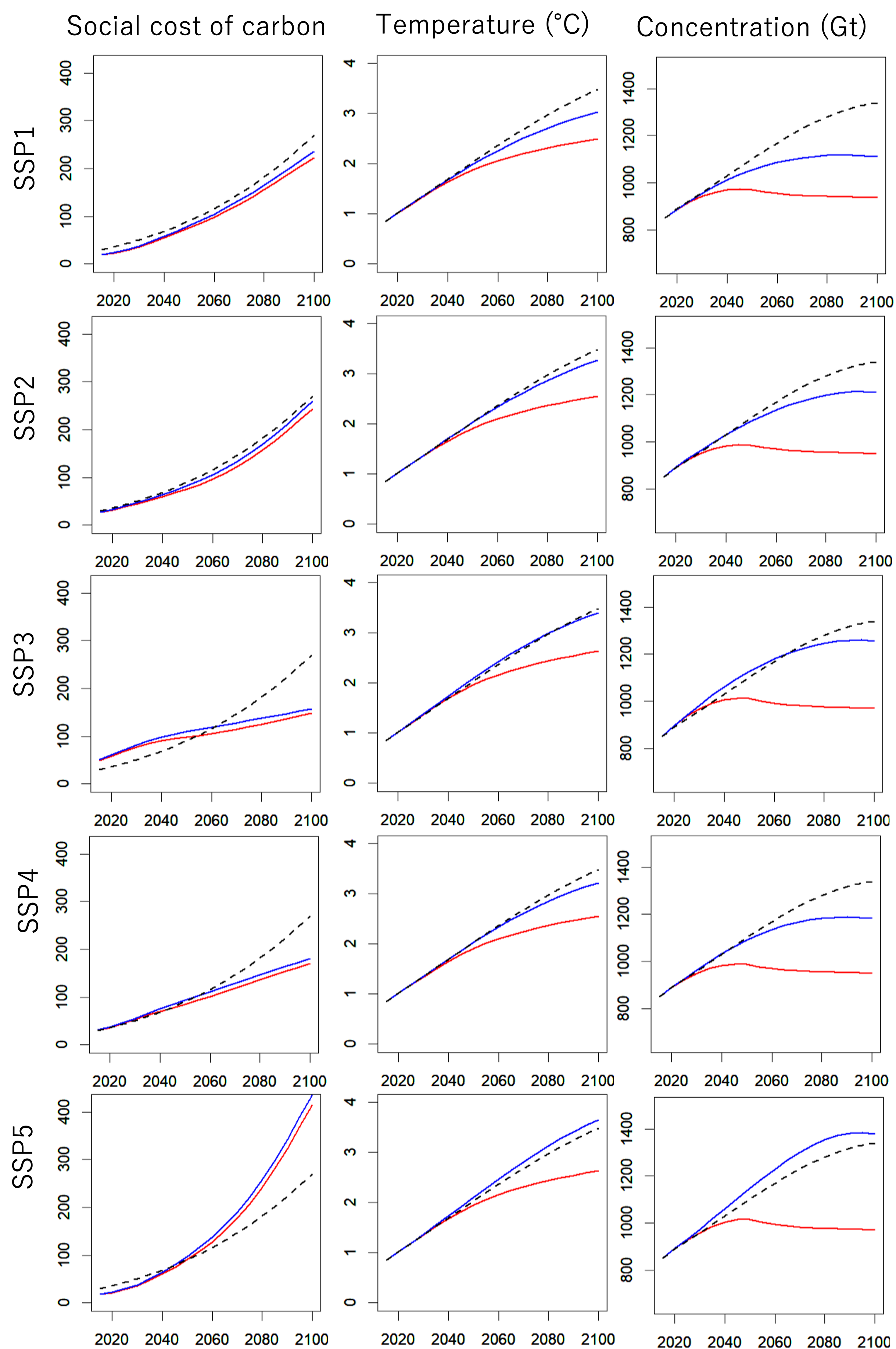}
  \caption{DICE results for social cost of carbon ($SCC_t$,  USD per ton of $\mathrm{CO}_2$), temperature ($T_t^{\mathrm{AT}}$), and carbon concentration ($M_t^{\mathrm{AT}}$) under five marker SSPs  under the scenarios achieving net-zero emissions in 2050 (red) and 2100 (blue). Black dashed line denotes the original DICE model results.}\label{fig_res_marker}
\end{figure}

Figure \ref{fig_res_marker} shows the DICE results for the trajectories of atmospheric temperature $T_t^{\mathrm{AT}}$, atmospheric carbon concentration $M_t^{\mathrm{AT}}$, and social cost of carbon $SCC_t$  (USD per ton of $\mathrm{CO}_2$) under \textbf{five SSP markers} from 2015 to 2100. Each subplot in Figure \ref{fig_res_marker}  shows the trajectory under the original DICE (black dashed line), the trajectory under  the scenario of achieving net-zero emissions in 2050 (red line), and the trajectory under the scenario of achieving net-zero emission in 2100 (blue line). Here, the original DICE trajectory is for reference. The original DICE trajectory corresponds to the optimal emission control and optimal consumption, and does not involve any SSP-IAM data calibration. A priori we do not know where the original DICE trajectory should be with respect to other scenarios plotted in Figure \ref{fig_res_marker}.

As expected we can see that the temperature under the scenario of net-zero in 2050 is lower than under the scenario of net-zero in 2100 across all SSPs. Under the 2050 net-zero scenario, the temperature in 2100 is about 2.5$^\circ$C for all SSPs while under the 2100 net-zero scenario, temperature in 2100 is just under 3$^\circ$C for SSP1 and materially larger than 3$^\circ$C for SSP2-SSP5 (highest under SSP5). This is of course not surprising because SSP1 corresponds to ``Green road" scenario and SSP5 is ``Fossil-fueled development" scenario. The original DICE trajectory of temperature is close 2100 net-zero scenario. This is somewhat expected because the optimal $\mu_t$ under original DICE reaches 1 in 2115 onwards.

Consistently with this behaviour of temperature, the carbon concentration under the 2050 net-zero scenario is always significantly lower than under the 2100 net-zero scenario. Concentration under 2050 net-zero scenario does not vary much across SSPs, while under the 2100 net-zero scenario concentration is the highest under SSP5 and the lowest under SSP1.

Another observation from Figure \ref{fig_res_marker} is that SCC trajectories are very different across SSPs but almost the same between the 2050 and 2100 net-zero scenarios.
Under SSP5, SCC is the largest. In 2100, it is approximately USD 400 under SSP5, USD 150 under SSP3, USD 180 under SSP4,  USD 250 under SSP2, and USD 200 under SSP1.
Note that SSP5 is the scenario of ``Fossil-fueled development" and thus it is not surprising to see it is leading to the largest social cost of carbon. It is interesting to note that the lowest SCC is under SSP3 where it is approximately USD 150 per ton in 2100 which is lower than under ``Green road" SSP1 scenario. Careful calculation/consideration is required for $SCC_t$ as it represents the net present value of all economic damages along the future trajectory from the additional ton of $\mathrm{CO}_2$ released in year $t$.
In 2025, SCC is approximately USD 30-50 across SSP markers and close to the original DICE results.
These results for SCC under SSP markers are generally consistent with reported results in the literature, see e.g. \cite{riahi2022mitigation} and  \citet{yang2018social}.

\newgeometry{top=0.25cm,left=1cm,right=1cm}
\restoregeometry

In addition to plots,
the temperature results under the marker SSPs for the year 2100 are shown in Table \ref{temperature_table_marker}.
Here, we show results for the net-zero scenarios and zero industrial emission scenarios. There is almost no difference for temperature between net-zero and zero industrial emission scenarios. Repeating what we observed in Figure \ref{fig_res_marker}, the smallest temperature 2.5$^\circ$C  is achieved  under SSP1 and net-zero in 2050, and it is close to 2.6$^\circ$C under other marker SSPs for net zero in 2050.
For the 2100 net-zero scenario, SSPs lead to more variation in temperature ranging from 3$^\circ$C under SSP1 to 3.7$^\circ$C under SSP5.

One can also observe that zero industrial emission scenario under marker SSP1 leads to slightly smaller temperature compared to the net-zero scenario. This is because the land-use emission is negative from 2050 under marker SSP1 (i.e. under SSP1-IMAGE) while positive under other marker SSPs. Negative land emission leads to negative total emission for zero industrial emission scenario, while net-zero scenario leads to only zero total emission resulting in the slightly higher temperature under the net-zero scenario.

\begin{table}[h!]
\captionsetup{width=.8\linewidth,skip=0pt}
\caption{Atmospheric temperature $T^\mathrm{AT}$ in 2100 under net-zero  and zero industry emission mitigation scenarios for marker SSPs.}\label{temperature_table_marker}
\begin{center}
\begin{tabular}{llllll}
\hline
& SSP1 & SSP2 & SSP3 & SSP4 & SSP5 \\
\hline
Net-zero in 2050 & 2.50 & 2.55 & 2.63 & 2.55 & 2.63 \\
Net-zero in 2100 & 3.03 & 3.26 & 3.40 & 3.22 & 3.65 \\
Zero industrial emission in 2050 & 2.49 & 2.62 & 2.75 & 2.63 & 2.68 \\
Zero industrial emission in 2100 & 3.02 & 3.29 & 3.45 & 3.25 & 3.66 \\
\hline
\end{tabular}
\end{center}
\end{table}


For non-marker scenarios, we consider four IAMs: AIM/CGE, GCAM, IMAGE and WITCH-GLOBIOM that have results under all five SSPs in the SSPdata-IIASA database. Table  \ref{temperature_table} shows atmospheric temperature in the year 2100 produced by the DICE model calibrated to the non-marker scenarios.
Here, we see results similar to Table \ref{temperature_table_marker}. Net-zero and zero industry emission scenarios lead to approximately the same results. Under the 2050 net-zero scenario, temperature range is 2.5-2.7$^\circ$C across SSPs and IAMs. Under the 2100 net-zero scenario, temperature range is 3.0-3.7$^\circ$C.  We also calculated trajectories of atmospheric temperature, atmospheric concentration and social cost of carbon under non-marker scenarios from 2015 to
2100 presented in Appendix A, Figures \ref{fig_res_AIM}-\ref{fig_res_WITCH-GLOBIOM}. These results are not materially different from the DICE results under the marker scenarios in Figure \ref{fig_res_marker}, confirming the consistency in predictions across considered process-based IAMs. Unfortunately, for all scenarios considered, we see that even under 2050 net-zero commitments, projected temperatures are in the range 2.5 -- 2.7$^\circ$C by 2100 exceeding 1.5 -- 2$^\circ$C target set out in the Paris Agreement, underscoring the need for negative emission technologies to be employed to achieve the target.

\begin{table}[H]
\captionsetup{width=.9\linewidth,skip=0pt}
\caption{Atmospheric temperature $T^\mathrm{AT}$ in 2100 under net-zero and zero industry emission mitigation scenarios for non-marker scenarios: five SSPs under four IAMs (AIM/CGE, GCAM, IMAGE, WITCH-GLOBIOM).}\label{temperature_table}
\begin{center}
\begin{tabular}{lllll|lllll}
\hline
& \multicolumn{4}{l}{Net-zero in 2050} & &\multicolumn{4}{l}{Zero industrial emission in 2050}  \\
SSP & AIM & GCAM & IMAGE & WITCH &  & AIM & GCAM & IMAGE & WITCH \\
1 & 2.5 & 2.5 & 2.5 & 2.5 &  & 2.5 & 2.4 & 2.5 & 2.5 \\
2 & 2.6 & 2.6 & 2.6 & 2.6 &  & 2.7 & 2.6 & 2.7 & 2.6 \\
3 & 2.6 & 2.6 & 2.6 & 2.6 &  & 2.8 & 2.8 & 2.8 & 2.7 \\
4 & 2.6 & 2.6 & 2.5 & 2.5 &  & 2.7 & 2.6 & 2.6 & 2.5 \\
5 & 2.6 & 2.6 & 2.7 & 2.6 &  & 2.7 & 2.6 & 2.7 & 2.7 \\
 &  &  &  &  &  &  &  &  &  \\
& \multicolumn{4}{l}{Net-zero in 2100} & &\multicolumn{4}{l}{ Zero industrial emission in 2100} \\
SSP & AIM & GCAM & IMAGE & WITCH &  & AIM & GCAM & IMAGE & WITCH \\
1 & 3.0 & 3.1 & 3.0 & 3.1 &  & 3.0 & 3.0 & 3.0 & 3.1 \\
2 & 3.3 & 3.3 & 3.3 & 3.3 &  & 3.4 & 3.3 & 3.3 & 3.3 \\
3 & 3.4 & 3.4 & 3.3 & 3.4 &  & 3.5 & 3.5 & 3.4 & 3.4 \\
4 & 3.1 & 3.2 & 3.2 & 3.2 &  & 3.2 & 3.3 & 3.2 & 3.2 \\
5 & 3.5 & 3.6 & 3.6 & 3.6 &  & 3.5 & 3.6 & 3.7 & 3.6 \\
\hline
\end{tabular}
\end{center}
\end{table}



\subsection{Stress testing life insurance and annuity portfolios}
In this subsection we use the mortality damage function derived in \cite{bressler2021mortality} in conjunction with the Gompertz law of mortality as a baseline mortality law.
We recall that the mortality damage function of \cite{bressler2021mortality} is of the shape $\widetilde\delta(t) = a t^\nu$ for $a = 0.0001811$ and $\nu = 3.745$.
Moreover, the Gompertz law of mortality for a 25-year old takes the form $\lambda(t) = \frac{1}{b}\exp\left((25+t-M)/b\right)$, where the parameters $M$ = 88.23 and $b = 9.38$ have been calibrated in \cite{arandjelovic23loads} to mortality rates from many countries; for details, we refer to the appendix therein.

\newpage

\begin{figure}[h!]
\centering
\captionsetup{width=.9\textwidth,format=plain}
\includegraphics[width=0.90\linewidth]{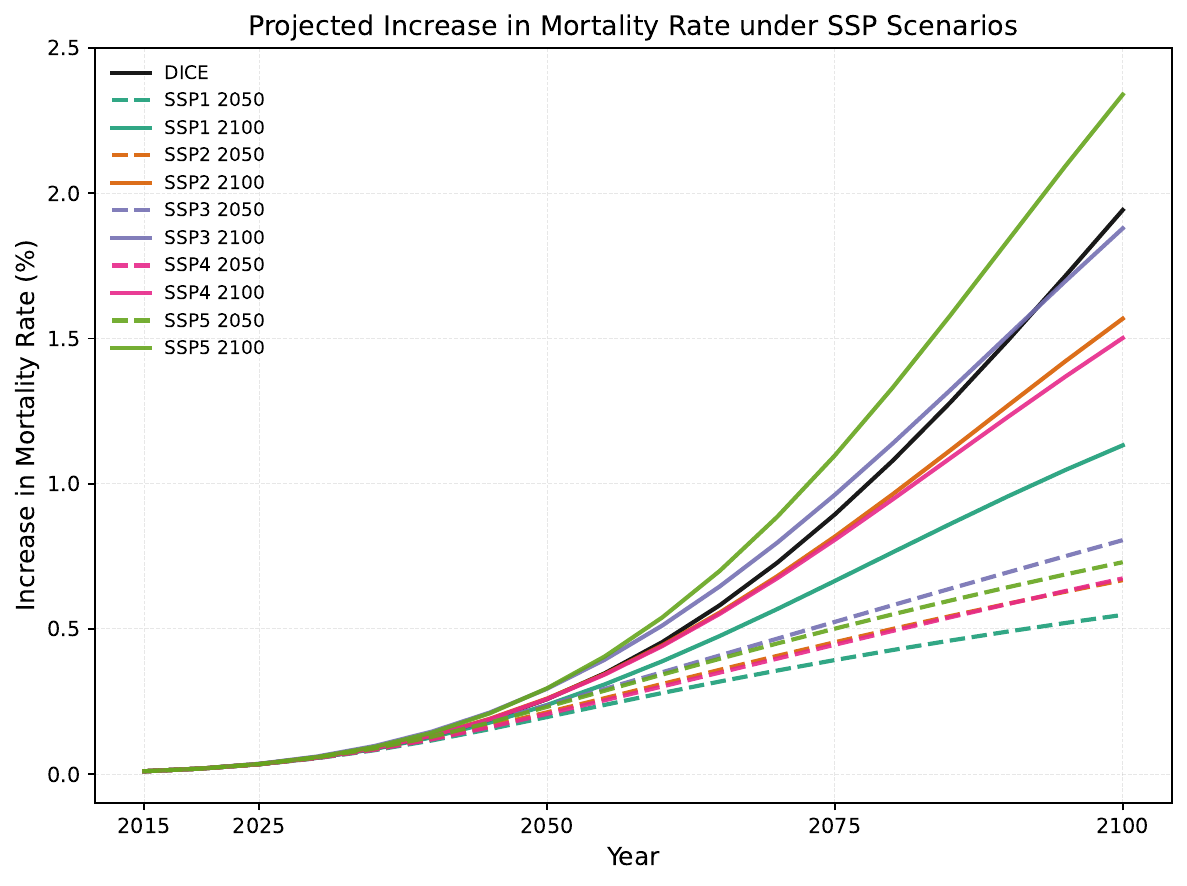}
\caption{Projected percentage increase in climate-induced excess mortality rates under the DICE-calibrated marker SSPs and net zero scenarios. Each curve represents the mortality impact implied by the corresponding temperature trajectory between 2015 and 2100.}
\label{fig:mortality-projection}
\end{figure}

Figure~\ref{fig:mortality-projection} demonstrates the evolution of projected climate-induced excess mortality under different marker SSP- and net-zero scenarios, as well as under the original DICE model.
We can observe a significant difference between the net zero 2050 and net zero 2100 policies.
The average net zero 2050 policy sees a mortality damage of around 0.6$\%$ in the year 2100, while the average net zero 2050 policy sees climate-attributed excess mortality of around 1.7$\%$ in the year 2100.
Excess mortality induced under the DICE trajectory becomes second-to-worst by 2100 with almost 2$\%$.

\begin{figure}[htbp]
\centering
\captionsetup{width=0.90\textwidth,format=plain}
\includegraphics[width=1.05\linewidth]{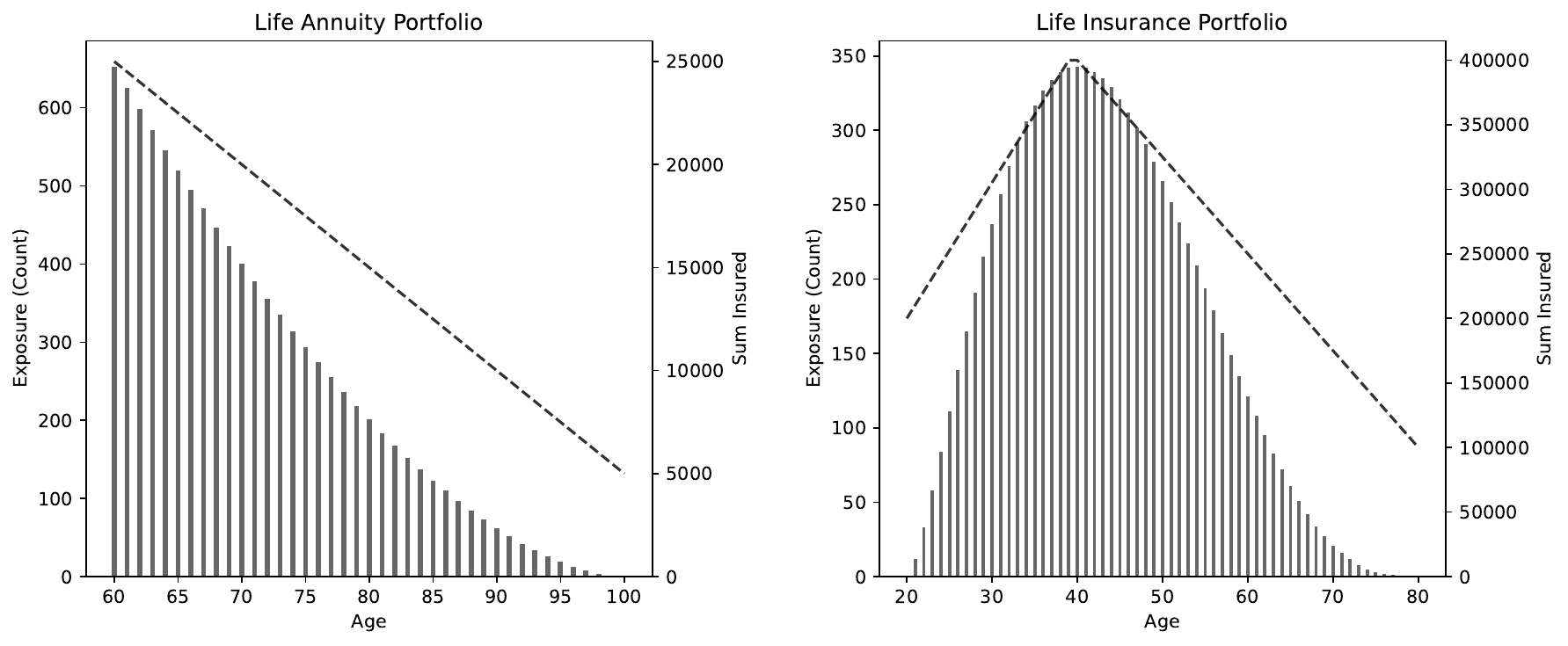}
\caption{Age distributions and corresponding total sum insured for the annuity and insurance portfolios. Bars show the number of clients at each age, while the dashed lines indicate the dollar value sum insured per client.}
\label{fig:age-distribution}
\end{figure}

For numerical illustration of the impact of temperature-related excess mortality rate on life-contingent insurance products we consider a projected stress test scenario for two portfolios in the year 2100 with age distributions and amounts insured similar to \cite{dong2024assessing} and \cite{tan2019comparing}, see Figure \ref{fig:age-distribution}:
\begin{itemize}
\item Portfolio A: 10,000 life annuity products,
\item Portfolio B: 10,000 life insurance products.
\end{itemize}
For portfolio A, we assume that the insured amount is paid out at the end of the year, given that the client survives until then.
We are interested in the distribution of the amount that is not paid out due to the fact that clients did not survive until the end of the year.
Climate-induced excess mortality on this distribution is expected to increase the amount not paid due to the increase in mortality rate.
Indeed, Table~\ref{tab:portfolio_scenarios} confirms this rationale an demonstrates moderate but significant increase in the amounts not paid, both for the mean amount as well as in the tails of the distribution (1\% and 99\% percentiles).
Similarly for portfolio B, where we are interested in the distribution of the total amount of insured sums paid out at the end of the year for every recorded client death, we can observe a noticeable increase in the mean amount and on the tails of the distribution.

~
~

\newpage

\begin{table}[h!]
\centering
\begin{threeparttable}
\caption{Life annuity and life insurance portfolios under DICE-calibrated SSP scenarios.}
\label{tab:portfolio_scenarios}
\small
\begin{tabular}{lrrrrrr}
\toprule
\multirow{2}{*}{\textbf{Scenario}} &
\multicolumn{3}{c}{\textbf{Annuity Portfolio}} &
\multicolumn{3}{c}{\textbf{Insurance Portfolio}} \\
\cmidrule(lr){2-4} \cmidrule(lr){5-7}
 & Mean & $q_{0.01}$ & $q_{0.99}$ & Mean & $q_{0.01}$ & $q_{0.99}$ \\
\midrule
DICE        & 1.96 & 2.13 & 1.86 & 1.95 & 2.81 & 1.63 \\
SSP1 2050   & 0.54 & 0.62 & 0.56 & 0.54 & 0.78 & 0.44 \\
SSP1 2100   & 1.14 & 1.23 & 1.10 & 1.13 & 1.79 & 0.93 \\
SSP2 2050   & 0.66 & 0.75 & 0.65 & 0.66 & 0.96 & 0.52 \\
SSP2 2100   & 1.58 & 1.75 & 1.51 & 1.58 & 2.25 & 1.31 \\
SSP3 2050   & 0.80 & 0.88 & 0.76 & 0.79 & 1.18 & 0.66 \\
SSP3 2100   & 1.90 & 2.05 & 1.81 & 1.89 & 2.57 & 1.54 \\
SSP4 2050   & 0.67 & 0.75 & 0.65 & 0.66 & 0.99 & 0.55 \\
SSP4 2100   & 1.52 & 1.67 & 1.45 & 1.51 & 2.04 & 1.26 \\
SSP5 2050   & 0.72 & 0.81 & 0.70 & 0.72 & 0.99 & 0.63 \\
SSP5 2100   & 2.38 & 2.61 & 2.26 & 2.37 & 3.16 & 1.99 \\
\bottomrule
\end{tabular}
\begin{tablenotes}[para]\footnotesize
Values represent relative deviations (\%) from the base case (no climate-induced impact on excess mortality) in mean, 1\% percentile ($q_{0.01}$), and 99\% percentile ($q_{0.99}$).
\end{tablenotes}
\end{threeparttable}
\end{table}

Finally, we check the impact of climate-induced excess mortality on human capital. We take the age-income function $y(s)$ from \cite{arandjelovic23loads} and calculate the human capital of a 25-year old in the year 2015, defined as the net present value of future income,
\begin{equation}
H = \int_0^T y(s) \mathrm{e}^{-rs-\int_0^s \widetilde{\lambda}(u)\, \mathrm{d}u}\, \mathrm{d}s,
\end{equation}
where $T=85$ denotes the maximal remaining lifetime, and $\widetilde{\lambda}(u) = \lambda(u) \times (1 + f(u))$, with $f(u)$ being the scenario-dependent mortality damage function as in Figure~\ref{fig:mortality-projection}, which has been approximated by a cubic polynomial for analytic tractability.
Table~\ref{tab:capital_scenarios} shows results.
We can observe an almost negligible impact of climate-induced mortality on human capital, which can be explained as follows. As evident from Figure~\ref{fig:mortality-projection}, the difference between projected mortality damages is almost negligible in the first half of this century, and only displays a more pronounced variability in the second half of the century.
Due to discounting ($r=3.2\%$), the moderate impacts on mortality become almost negligible when expressed in terms of net present value.

\begin{table}[h!]
\captionsetup{width=.7\linewidth,skip=0pt}
\caption{Human capital under DICE-calibrated marker SSP and net-zero scenarios. Values for $H$ are quoted in millions. $H_{\mathrm{rel}}$ denotes the relative deviation of $H$ from the base case in $\%$.}\label{tab:capital_scenarios}
\begin{center}
\small
\begin{tabular}{lrr}
\toprule
\textbf{Scenario} & $H$ & $H_{\mathrm{rel}}$ \\
\midrule
Base        & 1.4316 & - \\
DICE        & 1.4314 & -0.01 \\
SSP1 2050   & 1.4315 & -0.01 \\
SSP1 2100   & 1.4314 & -0.01 \\
SSP2 2050   & 1.4315 & -0.01 \\
SSP2 2100   & 1.4314 & -0.01 \\
SSP3 2050   & 1.4314 & -0.01 \\
SSP3 2100   & 1.4314 & -0.02 \\
SSP4 2050   & 1.4315 & -0.01 \\
SSP4 2100   & 1.4314 & -0.01 \\
SSP5 2050   & 1.4315 & -0.01 \\
SSP5 2100   & 1.4313 & -0.02 \\
\bottomrule
\end{tabular}
\end{center}
\end{table}

\section{Conclusions}
\label{sec:conclusion}
In this paper we developed climate-economy scenario trajectories using the cost-benefit DICE model calibrated to five baseline SSPs and six main process-based IAMs. For emission mitigation control, we considered two scenarios: achieving net-zero in 2050 and in 2100. We also considered scenarios of achieving zero industrial emission in 2050 and in 2100. We showed how the excess temperature-related mortality rate can be calculated for these projections, providing stress scenarios for life insurer portfolios. These scenarios of global temperature can be used for non-life insurance scenarios for events related to rising temperature. In this case, one needs to get estimates of frequency and severity of such insurable events as a function of global temperature that will directly translate to the quantifiable impact on non-life insurance products. We did not consider this and leave this to further research.

One of the main observations for the obtained climate-economy scenarios is that even if net-zero emission (or zero industrial emission) is achieved by 2050, the temperature will exceed $2^\circ\text{C}$ by 2100 (ranging from $2.5^\circ\text{C}$ to $2.7^\circ\text{C}$) under all five SSPs and six IAMs considered. For more lenient mitigation achieving zero emission by 2100, the temperature will be in the range between $3^\circ\text{C}$ and $3.7^\circ\text{C}$ by 2100 depending on the SSP-IAM. The social cost of carbon is raising from USD 30-50 in 2025 to USD 250-400 in 2100 depending on SSP-IAM with only very marginal impact from net-zero scenarios. This is generally consistent with other literature, see e.g. \cite{riahi2022mitigation} and  \citet{yang2018social} and provides a carbon price benchmark for policy makers.
SCC does not  depend significantly on the emission mitigation scenario because it measures the marginal damage to the economy from carbon emissions.

As expected, the smallest temperature corresponds to SSP1 and largest to SSP5. Here, we note, that temperature trajectories in \cite{yang2018social} are in the range 3.75-4.5 $^\circ$C in 2100 across five SSPs higher than in our study because they are calculated for optimal emission reduction $\mu_t$ while our results correspond to scenarios for $\mu_t$ leading to zero emissions in 2050 and 2100.
The results are aligned with the findings in \cite{tol2023costs} that the Paris Agreement temperature targets cannot be met without very rapid reduction
of greenhouse gas emissions and removal of carbon dioxide from the atmosphere using negative emission technologies \cite[][]{santos2019negative}. The latter
requires perhaps prohibitively large subsidies.

Various sources of uncertainty can be considered in the DICE model, such as uncertainties in the damage function, total productivity growth, or decarbonisation function decline rate that have a significant impact on the future trajectories of climate and economy not considered in our study. These have been studied as parameter uncertainties in \cite{nordhaus2018projections}.
Stochastic versions of DICE where optimal policy is calculated as a decision under uncertainty solving optimal stochastic control problem are also considered in \cite{traeger2014fourstatedDICE, cai2019social, arandjelovic2024solving,shevchenko2022impact}.
Considering stochastic DICE models with exogenous functions calibrated to SSPs should also provide important benchmarks and will be considered in our future studies.

Finally, it is important to note that even though DICE model was introduced three decades ago it is regularly updated with the most recent beta release in 2023. Also, there is a recent study \cite{folini2025climate} finding that the functional form of climate equations in DICE 2016 is fit for purpose, but parameters of these equations should be recalibrated. These updates should be considered in practical implementation of stress test assessment proposed in our paper.

\section*{Acknowledgements} Pavel Shevchenko and Tor Myrvoll acknowledge travel support from the Institute of Statistical Mathematics, Japan. We also thank the participants of the Workshop ``Advances in Risk Modelling and Applications to Finance and Climate Risk" held in July 2024 in Vienna and ``Climate Finance \& Risk 2024" in November 2024 in Japan for their helpful comments on our preliminary results. This work was supported by JSPS KAKENHI Grant Number 21K18309.


\paragraph{Funding Statement}
This work was partially supported by JSPS KAKENHI Grant Number 21K18309.

\paragraph{Competing Interests}
All authors declare that they have no competing interests.




\newpage

\bibliographystyle{elsarticle-harv}
\bibliography{Bibliography}

\newpage

\newgeometry{top=0.5cm,left=1cm,right=1cm, bottom=0.25cm}
\appendix
\section{Appendix: Social cost of carbon, temperature, and concentration under various non-marker SSPs}

\begin{figure}[!h]
  \centering
  \includegraphics[width=15cm]
  {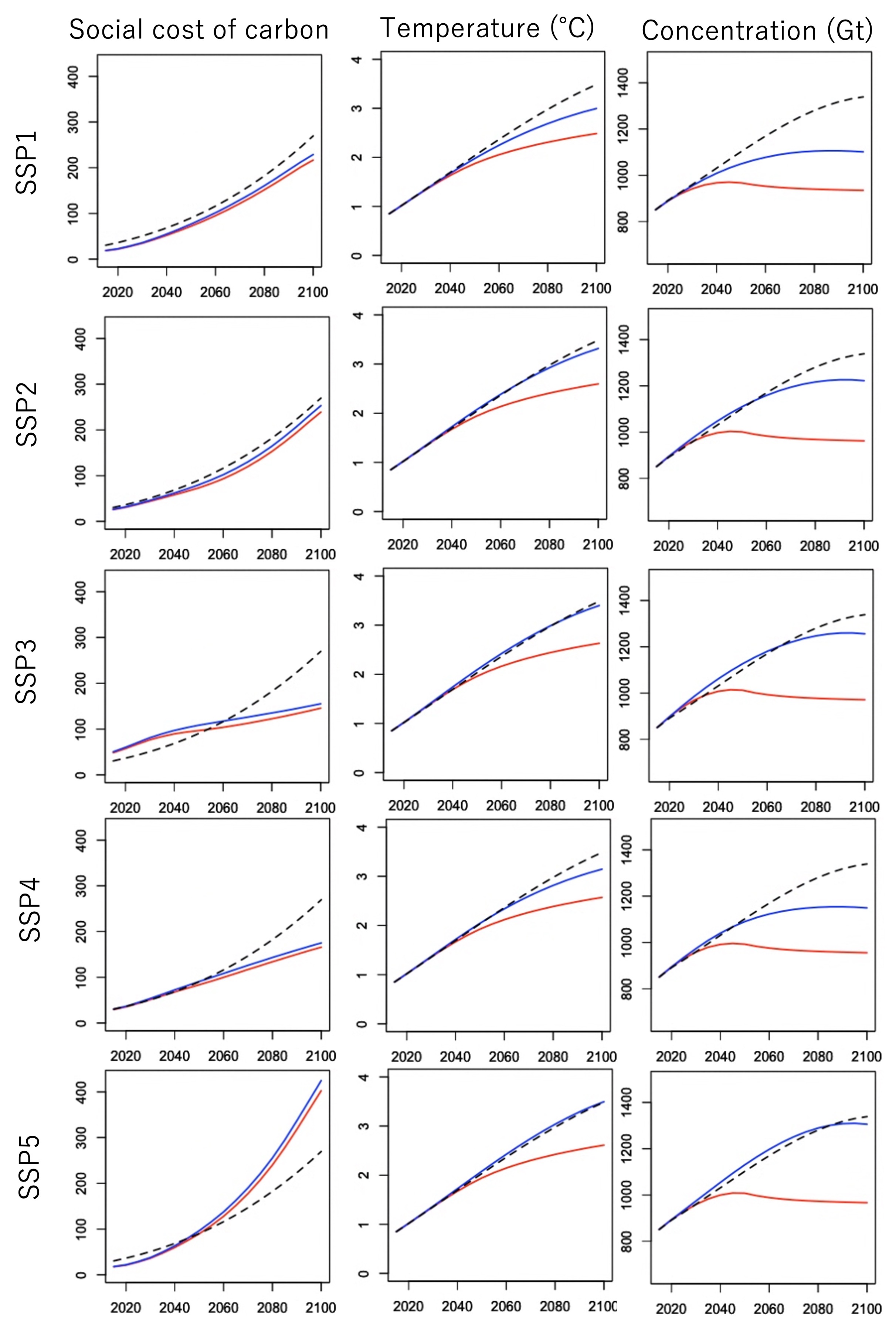}
  \caption{Social cost of carbon ($SCC_t$,  USD per ton of $\mathrm{CO}_2$), temperature ($T_t^{\mathrm{AT}}$), and carbon concentration ($M_t^{\mathrm{AT}}$) from the DICE model calibrated to baseline SSPs in \textbf{AIM} for scenarios of achieving net-zero emissions in 2050 (red) and 2100 (blue). Black dashed line denotes the original DICE model results.}\label{fig_res_AIM}
\end{figure}
\restoregeometry

\newgeometry{top=0.25cm,left=1cm,right=1cm}
\begin{figure}
  \centering
  \includegraphics[width=15cm]{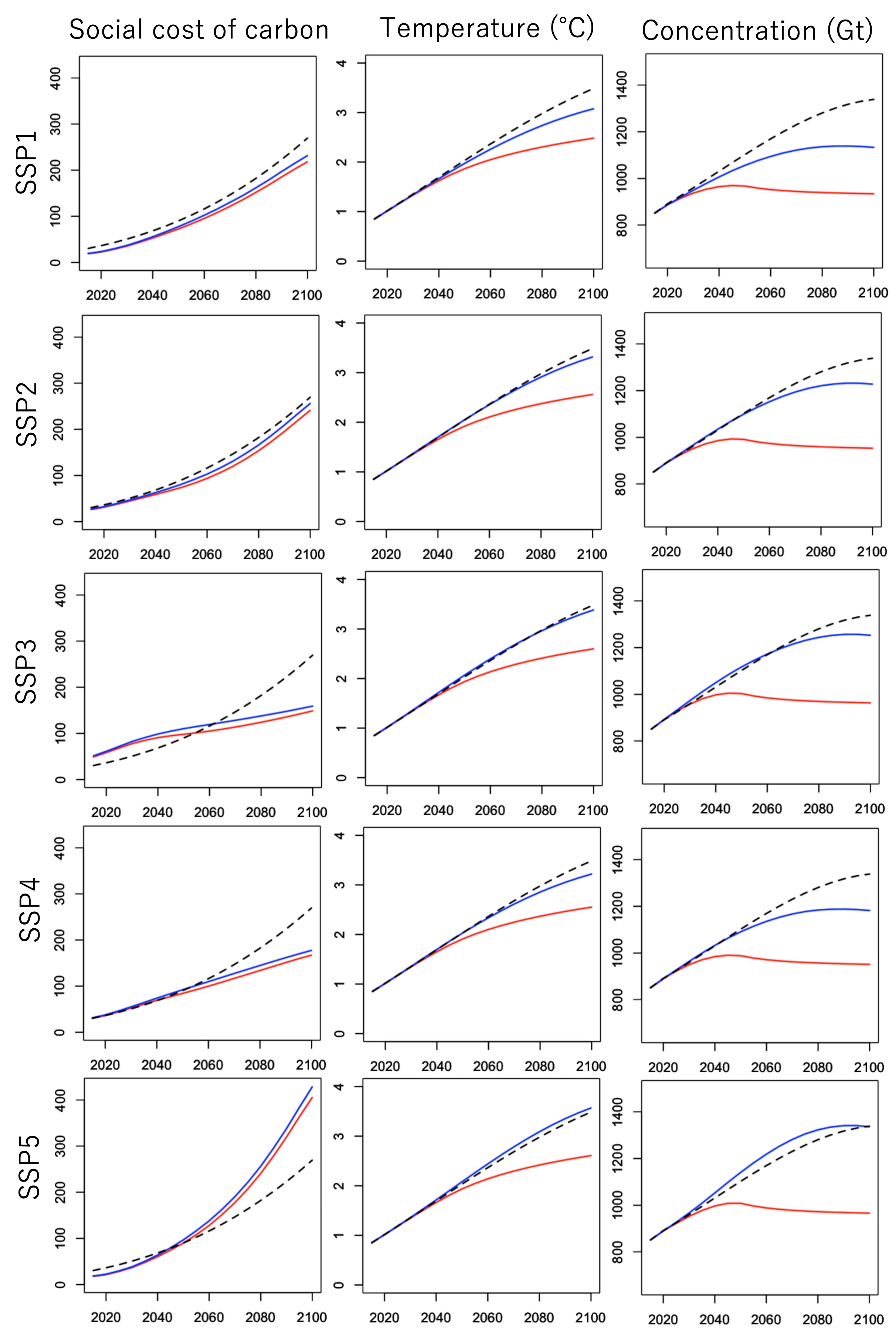}
  \caption{Social cost of carbon ($SCC_t$, USD per ton of $\mathrm{CO}_2$), temperature ($T_t^{\mathrm{AT}}$), and carbon concentration ($M_t^{\mathrm{AT}}$) from the DICE model calibrated to baseline SSPs in \textbf{GCAM} for scenarios of achieving net-zero emissions in 2050 (red) and 2100 (blue). Black dashed line denotes the original DICE model results.}\label{fig_res_GCAM}
\end{figure}

\begin{figure}
  \centering
  \includegraphics[width=15cm]{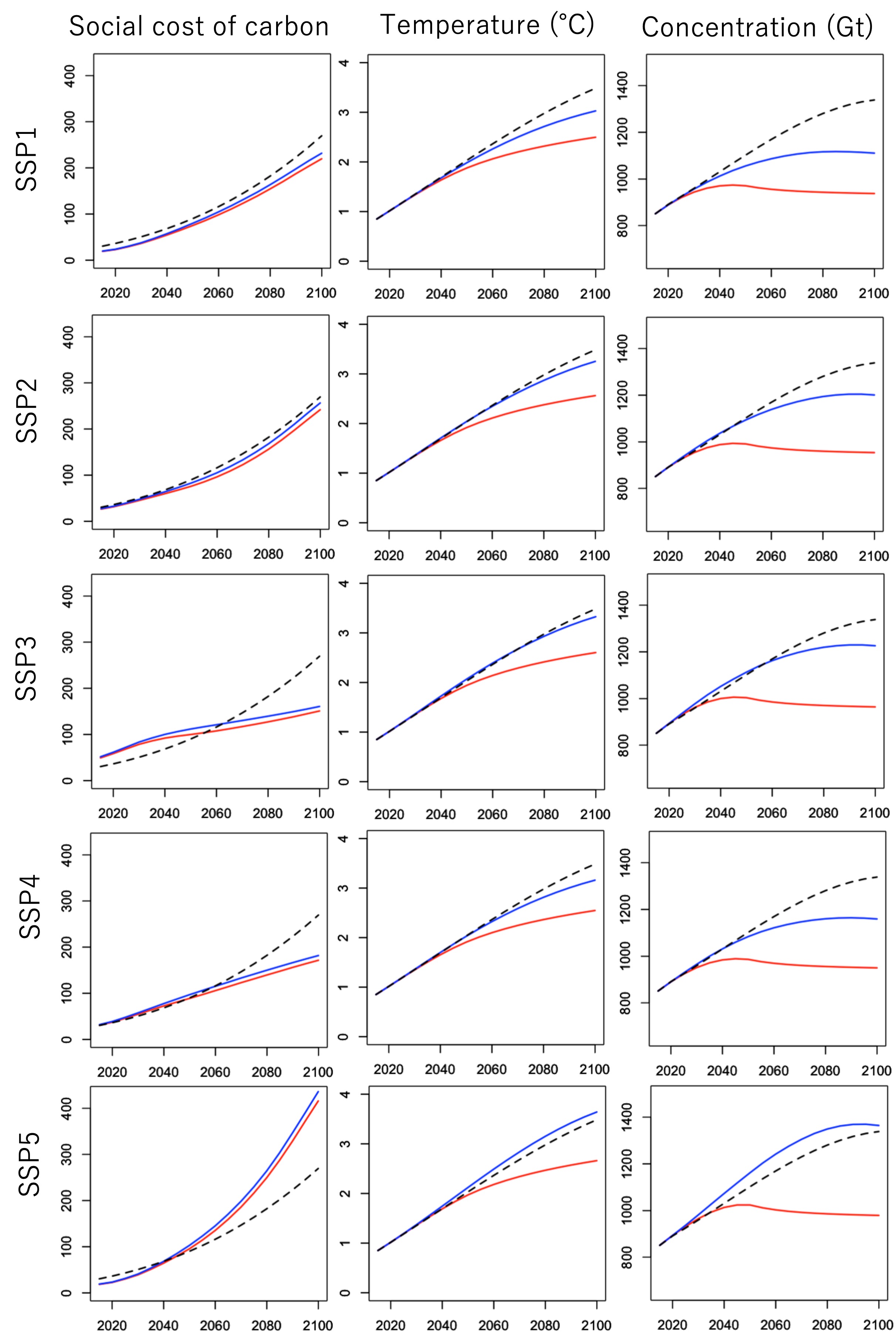}
  \caption{Social cost of carbon ($SCC_t$,  USD per ton of $\mathrm{CO}_2$), temperature ($T_t^{\mathrm{AT}}$), and carbon concentration ($M_t^{\mathrm{AT}}$) from the DICE model calibrated to baseline SSPs in \textbf{IMAGE} for scenarios of achieving zero emissions in 2050 (red) and 2100 (blue). Black dashed line denotes the original DICE model results.}\label{fig_res_IAMGE}
\end{figure}

\begin{figure}
  \centering
  \includegraphics[width=15cm]{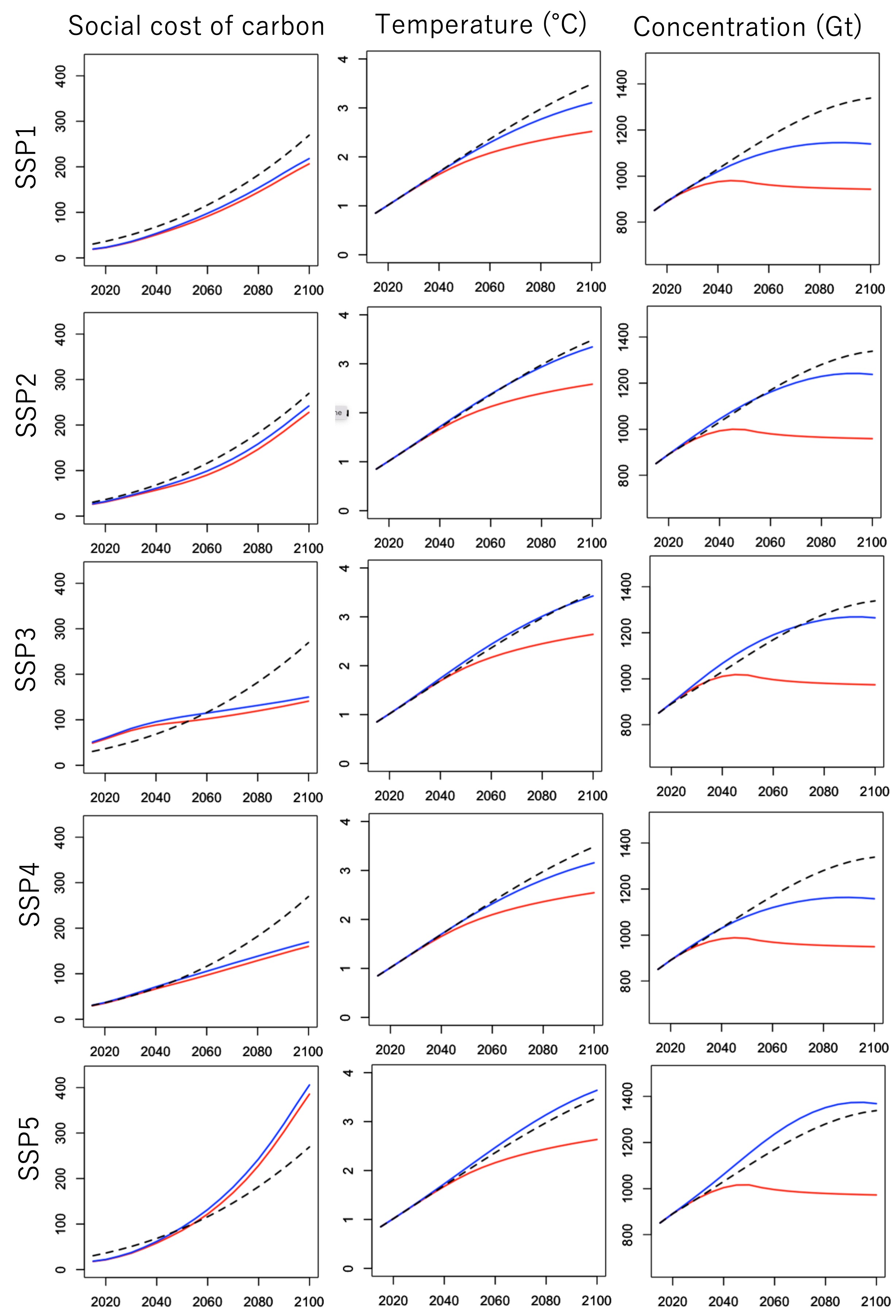}
  \caption{Social cost of carbon ($SCC_t$,  USD per ton of $\mathrm{CO}_2$), temperature ($T_t^{\mathrm{AT}}$), and carbon concentration ($M_t^{\mathrm{AT}}$)  from the DICE model calibrated to baseline SSPs in \textbf{WITCH-GLOBIOM} for scenarios of achieving net-zero emissions in 2050 (red) and 2100 (blue). Black dashed line denotes the original DICE model results.}\label{fig_res_WITCH-GLOBIOM}
\end{figure}
\restoregeometry

\end{document}